# Structure and function of Enterotoxigenic *Escherichia coli* fimbriae from differing assembly pathways

Running title: Structure and Function of Two ETEC Fimbriae

Narges Mortezaei[a], Chelsea R. Epler[b], Paul P. Shao[b], Mariam Shirdel[a], Bhupender Singh[a,c], Annette McVeigh[d], Bernt Eric Uhlin[c], Stephen J. Savarino[d,e], Magnus Andersson[a*], and Esther Bullitt[b*]

[a]Department of Physics, Umeå University, SE-901 87 Umeå, Sweden, [b]Department of Physiology and Biophysics, Boston University School of Medicine, Boston, MA 02118, USA, [c]The Laboratory for Molecular Infection Medicine Sweden (MIMS) and Department of Molecular Biology, Umeå University, SE-901 87 Umeå, Sweden, [d]Enteric Diseases Department, Infectious Diseases Directorate, Naval Medical Research Center, Silver Spring, MD 20910, USA, and [e]Department of Pediatrics, Uniformed Services University of the Health Sciences, Bethesda, MD 20814, USA.

* Corresponding authors
Email: bullitt@bu.edu

## Abstract

Pathogenic enterotoxigenic *Escherichia coli* (ETEC) are the major bacterial cause of diarrhea in young children in developing countries and in travelers, causing significant mortality in children. Adhesive fimbriae are a prime virulence factor for ETEC, initiating colonization of the small intestinal epithelium. Similar to other Gram-negative bacteria, ETEC express one or more diverse fimbriae, some assembled by the chaperone-usher pathway and others by the alternate chaperone pathway. Here we elucidate structural and biophysical aspects and adaptations of each fimbrial type to its respective host niche. CS20 fimbriae are compared to CFA/I fimbriae, which are two ETEC fimbriae assembled via different pathways, and to P-fimbriae from uropathogenic *E. coli*. Many fimbriae unwind from their native helical filament to an extended linear conformation under force, thereby sustaining adhesion by reducing load at the point of contact between the bacterium and the target cell. CFA/I fimbriae require the least force to unwind, followed by CS20 fimbriae and then P-fimbriae, which require the highest unwinding force. We conclude from our electron microscopy reconstructions, modeling, and force spectroscopy data that the target niche plays a central role in the biophysical properties of fimbriae that are critical for bacterial pathophysiology.

## Introduction

Diarrhea is the second leading cause of death in children under 5 years of age (Liu *et al.*, 2010). Among the most common causes of diarrhea, enterotoxigenic *Escherichia coli* (ETEC) is estimated to be responsible for over 121,000 deaths per year (Liu *et al.*, 2010). ETEC is also a major causative agent of travelers' diarrhea (Lozano *et al.*, 2012). After ingestion via contaminated food or water, ETEC transit the stomach into the small intestine, where they colonize epithelial tissue. Colonization is initiated by the key step of adhesion, mediated by bacterial fimbriae. A number of human ETEC fimbrial colonization factors have been described, many of which fall into Class 1b or Class 5 fimbriae based on phylogenetic relatedness of their major pilin subunits (Low, 1996, Girardeau *et al.*, 2000). The former include CS12, CS18, CS20 and likely several additional members that have been recently distinguished (Tacket *et al.*, 1988, Honarvar *et al.*, 2003, Valvatne *et al.*, 2004, Nada *et al.*, 2011). CS20 was identified in 6% of a collection of ETEC isolates from Indian children with diarrhea (Valvatne *et al.*, 2004) and may also be found in other diarrheagenic pathotypes of *E. coli* (Ochoa *et al.*, 2010). CS20 and other Class 1b fimbriae are assembled by the chaperone-usher pathway (CUP), as are other fimbriae including Type 1 fimbriae and P-fimbriae expressed by uropathogenic *Escherichia coli* (UPEC). Certain sequence motifs are typical of CUP fimbriae, including alternating hydrophobic residues at the C-terminus of their pilins (Dodson *et al.*, 1993, Nishiyama *et al.*, 2005, Nishiyama *et al.*, 2008). The pilins of Class 5 fimbriae, of which ETEC CFA/I fimbria is the archetypal member, do not share this or other signature motifs with CUP fimbriae and are assembled by what has been termed the alternate chaperone pathway (ACP), using chaperones distinct from those of the CUP (Soto & Hultgren, 1999, Bao *et al.*, 2014)

In both the CU and AC pathways, pilins are secreted across the inner membrane through the general secretory pathway and are chaperoned across the periplasm. At the outer membrane, an usher protein facilitates donor-strand exchange, in which the N-terminus of each pilin subunit is transferred from the chaperone protein to a growing chain of pilin subunits (Verger *et al.*, 2007, Choudhury *et al.*, 1999, Sauer *et al.*, 1999). All aforementioned fimbriae assemble into helical structures 1-3μm in length, with an outer diameter of approximately 70-80 Å, and 3.0-3.3 subunits per turn of the helix (Mu *et al.*, 2008, Mu *et al.*, 2005). Despite these similarities, each of these fimbrial types has distinct target tissues, amino acid sequences, biomechanical properties, and structural features, all of which are specialized for survival in its respective host niche. Bacteria that express CS20 fimbriae, like those that express CFA/I, colonize the intestinal tract and cause diarrheal disease. In contrast, the preferred pathogenic niche of bacteria expressing P-fimbriae is the urinary tract. Genetically, major pilins from fimbriae of human pathogens within a pathway share 20-23% sequence identity, and pilin sequences between pathways have more limited identity, 15-16% (Table 1). That is, amino acid sequences of pilins from the CUP family differ by the same extent whether they are pathogens of the urinary tract or the intestines. Biomechanical properties of fimbriae facilitate sustained adhesion of bacteria, despite differing shear forces in the specific target microenvironment of each bacterial pathotype.

In this work, we report that CS20 pilins assemble into fimbriae that are straighter and less unwound than their CFA/I fimbrial counterparts. Moreover CS20 fimbriae require a higher force to unwind than CFA/I fimbriae but a lower force than UPEC P-fimbriae. The structure of CS20 fimbriae was determined by 3D reconstruction of electron cryomicroscopy data. Homology modeling and fitting of the major pilin subunit CsnA into the 3D map was used to propose that significant bonding forces hold the layers of the helix together. Our data provide evidence for pathophysiological adaptation of virulence fimbriae on bacteria colonizing the lower small intestine.

## Results

**Genetic determinants of CS20 fimbriae typical of Class 1b CUP fimbriae**

CS20 fimbriae have been classified as Class 1b fimbriae based on the reported sequence of its major subunit (Valvatne *et al.*, 2004). We determined the sequence of the entire CS20 fimbrial gene cluster from enterotoxigenic *E. coli* (ETEC) strain WS7179A-2 (Genbank Accession No. KJ922517), a strain originally isolated from a young child with diarrhea. As anticipated, the organization of the gene cluster (Fig. S1) was identical to that previously reported for CS12 and CS18, both colonization factor fimbriae of ETEC from humans, as well as the organization for 987P, a porcine ETEC fimbria (Honarvar *et al.*, 2003). WS7179A-2 expresses CS20 fimbriae that are recognized by fimbriae-specific polyclonal antibodies and show a positive adherence phenotype to Caco-2 cells (data not shown). With introduction of pRA101, a plasmid containing *cfaD*, an *araC*-like positive transcriptional regulator, CS20 expression is augmented and fimbria production is increased in ETEC WS7179A-2/pRA101.

**CS20 fimbriae are straighter and have fewer unwound regions than CFA/I and P fimbriae**

To confirm expression of CS20 fimbriae and investigate their morphology, we imaged ETEC WS7179A-2/pRA101 using atomic force microscopy (AFM). An example of a cell grown under normal conditions with fimbriae extending from the cell surface is shown in Fig. 1A. The approximate length of the fimbriae is 2 μm, and fimbriae were rarely found in an unwound (linearized) state, suggesting that the layer-to-layer bonds between subunits on adjacent turns of the helix are sufficiently strong to keep the quaternary structure intact when prepared without significant hydrodynamic forces. Thus, we expected that there were more subunit-subunit interactions between layers and that a higher force would be required to unwind CS20 in force spectroscopy experiments in comparison to CFA/I fimbriae (Andersson *et al.*, 2012). Upon further investigation by transmission electron microscopy (TEM), we found small unwound sections of fimbriae (Fig. 1B, yellow arrow) among mostly intact fimbriae (Fig. 1B, blue arrow) in samples that had been frozen in vitreous ice. Occasionally, short segments of fimbriae are observed end-on, providing a view of the hollow channel along the helical axis (Fig. 1B, red arrow). CS20 fimbriae (Fig. 1B) are significantly straighter than CFA/I fimbriae on ETEC or P-fimbriae on UPEC (Mu & Bullitt, 2006, Mu *et al.*, 2008) (Fig. S2A and S2B, respectively). To quantify the change in stiffness of these fimbriae, we calculated a persistence length for CS20 fimbriae; persistence length quantifies the stiffness of the structure, where a high number indicates a stiffer structure. The resulting persistence length of 28 µm is significantly longer than persistence lengths for CFA/I and P-fimbriae, 1.5 µm and 7 µm, respectively (Bullitt & Makowski, 1998, Andersson *et al.*, 2012), showing that CS20 fimbriae are stiffer than other similar fimbriae.

**Forces required to unwind CS20 fimbriae are intermediate between those needed to unwind CFA/I and P fimbriae**

To quantify the force required to extend and rewind ETEC CS20 fimbriae, bacteria expressing CS20 fimbriae were assessed by optical tweezers force spectroscopy. Measurements were performed on single CS20 fimbriae using a similar approach to that previously described (Andersson *et al.*, 2006b). A typical force response of a CS20 fimbria under steady-state conditions is shown in Fig. 2. The force response under extension shows a similar response to both the previously studied P and Type 1 adhesion fimbriae expressed by UPEC and CFA/I expressed by ETEC (Andersson *et al.*, 2008, Andersson *et al.*, 2012). Initially, the force increases linearly with extension, representing an elastic response of the shaft. Thereafter, the force is constant with extension, representing sequential unwinding of the helical shaft subunits. Finally, the force increases linearly, representing elastic stretching of the subunits. The contraction response tracks that of extension, except for a dip in force (at ~4 µm in Fig. 2) before rewinding under a constant force. The dip in force is similar to what has been seen for other helical fimbriae, e.g., P, Type 1, S, F1C, CFA/I fimbriae (Andersson *et al.*, 2008, Castelain *et al.*, 2011, Castelain *et al.*, 2010). That is, upon unwinding, and thereby a change from a helical to a linear configuration, subunit-subunit interactions are broken, and the subunits move apart. These subunits, which in the helical configuration are positioned in adjacent layers, must re-associate, and slack in the fimbria is needed for restoration of a helical filament.

To find the average unwinding force, which provides information on the strength of the layer-to-layer bonds, we averaged the force-extension data of region II. For all data curves (*n=58*) the average unwinding force was 15 ± 1 pN. This force is lower than for P and Type 1 fimbriae expressed by UPEC that unwound at 28 pN and 30 pN, respectively. However, it is higher than for ETEC CFA/I fimbriae, which unwound at 7.5 pN. In addition, similar to what has been noted for other helical fimbriae, and shown in Fig. S3, repetitive measurements on the same CS20 fimbria showed that force responses of each cycle were identical, i.e., fimbriae are not damaged during the extension or rewinding process.

**Dynamic properties of CS20 fimbriae**

To define dynamic properties of the CS20 fimbriae and complement our "standard" (steady-state) force-extension measurements, we performed dynamic force spectroscopy (DFS) (Andersson *et al.*, 2006a). DFS was applied to CS20 fimbriae to quantify both the corner velocity, $\dot{L}^*$, which defines the maximum extension velocity for a fimbria before a further increase in velocity requires an increase in the applied force, and the layer-to-layer bond opening length, $\Delta x_{AT}$. A single fimbria was first unwound in order to find the total length of region II. Thereafter, this fimbria was rewound and again partially unwound, typically a few micrometers, before the experiment was begun. Figure 3A shows an example of a DFS measurement demonstrating that unwinding force increases as a function of the applied velocity. Measurements were acquired at 5000 Hz sampling rate.

The average unwinding force for all fimbriae ($n = 13$) at a given unwinding velocity was thereafter plotted as a force versus extension velocity diagram and the two models presented in the supplementary information section were fitted to the data, as seen in Fig. 3B. The fit (blue dashed line) using Eq. (2), which neglects the refolding rate (Andersson *et al.*, 2006a), shows good agreement with the experimental data for high extension velocities. The intersect between the gray and blue dashed lines represent the corner velocity, defining the minimum speed at which increased force is needed to unwind the fimbrial rod, which was measured to be $\dot{L}^* = 877$ nm/s. For a more comprehensive quantitation, we also fitted these data with full rate equations, as described by Eq. (1). As can be seen in Fig. 3B, the model (red dashed line) fits the data well with resulting values of the variables: $F_{uf}^0 = 15$ pN, $\Delta x_{AT} = 0.4$ nm, and $\dot{L}^* = 824$ nm/s. Thus, the dynamic response of the CS20 fimbriae is predicted by the biophysical sticky-chain model, and is compared to other fimbriae in table S1.

**Three-dimensional reconstruction of CS20 fimbriae**

To better understand the details of the structure that provides these biomechanical properties, a helical reconstruction was carried out on CS20 fimbriae preserved in vitreous ice. First, STEM data from Brookhaven National Lab were collected and analyzed to define the mass per unit length along the fimbriae as 1973 ± 21 Da/Å. Using the known molecular weight of the pilin subunit CsnA, 17520 Da (Valvatne *et al.*, 2004), the rise per subunit of the helix was calculated to be 8.9 ± 0.1 Å.

Images of CS20 fimbriae were taken under frozen hydrated conditions and the three-dimensional reconstruction was computed using SPARX Helicon helical reconstruction software (Behrmann *et al.*, 2012, Hohn *et al.*, 2007). A solid cylinder of noise with an approximately correct diameter (80 Å) was used as an *ab initio* reference model. For the reconstruction 172,716 particles were selected from 2,787 filament segments with a 96% overlap between boxes (the spacing between boxes was slightly larger than the rise per subunit; 9.05 Å cf 8.9 Å). After iteration, the final reconstruction shown in Fig. 4B was calculated from the final class averages, and included 91% of the particles. We report here the resolution of CS20 fimbriae at 10.3 Å, as determined by the conservative 0.5 cutoff of the Fourier shell coefficient. We found that CS20 fimbriae have an outer diameter of 82 Å and the central channel has an inner diameter of 33.5 Å. CS20 fimbriae have helical symmetry of 3.21 subunits per turn of the helix, an 8.9 Å rise per subunit, and the rotation of subunits around the helical axis is 112.3º. The pitch of the helix is therefore 28.5 Å. Handedness of the fimbria was determined using rotary shadowed data, and a surface view of the final reconstruction shown in Fig. 4B reveals a left-handed long-pitch helix and a right-handed genetic (one-start) helix. The layer-to-layer interactions between subunits of CS20 (Fig. 5, green) appear to be less robust than in P-fimbriae (Fig. 5, pink) and more than are observed in CFA/I fimbriae (Fig. 5, blue), as expected from the force data above.

**CsnA has similar secondary structure to the Type 1 fimbriae subunit, FimA**

To define interactions between subunits in the CS20 structure, we aimed to fit CsnA, the major pilin monomer, into the EM map; however, currently there is no crystal structure of CsnA available. CS20, PapA, and Type 1 fimbriae are all members of the class 1 adhesion fimbriae family (Fig. S4), and the structures of PapA (PDB: 2UY6) and FimA (PDB: 2JTY) are known (Puorger *et al.*, 2011, Verger *et al.*, 2007, Dereeper *et al.*, 2008). The sequence of the mature CS20 major pilin CsnA from WS7179A-2 discussed here is identical to a previously published sequence (Valvatne *et al.*, 2004). When the amino acid sequences of FimA and CsnA were aligned using ClustalW2 (Larkin *et al.*, 2007) they showed 42% similarity and 23% identity, as compared to 42% similarity and 22% identity for PapA and CsnA (Table 1). After using MODELLER software (Sali *et al.*, 1995) to model CsnA from both known pilin structures, an initial model was calculated for CsnA using FimA as the reference due to its slightly more compact structure. The model for CsnA was energy minimized, and is shown in Fig. 4A (purple ribbon).

We further tested the validity of the homology modeling by investigating the surface charge of CsnA both computationally and experimentally. The hydrophobicity and surface potentials calculated using UCSF Chimera software (Pettersen *et al.*, 2004) indicated that the surface of this subunit was about equally hydrophobic and hydrophilic, as shown in Fig. S5A. The surface potential map of CsnA indicates that CsnA has a negative surface potential, as shown in Fig. S5B.

Trials to test attachment of fimbriated bacteria to either aldehyde/sulfate (negatively charged) or amidine (positively charged) beads were evaluated at different pH values with optical tweezers. We expected the best attachment of CS20 fimbriae to hydrophobic amidine beads, due to the positively charged bead surface and tested bacterial binding to beads in buffers at pH 6.80 and 7.40. In the experiments (*n*=15 for each combination) it was found that the success rate using the amidine beads at a pH of 6.80 was the highest, verifying the negatively charged surface of CsnA. The homology model could then be used to understand how the CsnA subunit may fit into the CS20 three-dimensional reconstruction and to elucidate interactions between subunits.

**CS20 is held together by donor-strand-exchange for subunit-subunit interactions and layer-to-layer bonding of CsnA monomers.**

To more clearly identify the subunit-subunit interactions in CS20 fimbriae, we docked the homology modeled CsnA into the CS20 EM map. A satisfactory fit was not possible with the homology modeled structure. Based upon the knowledge that other adhesion fimbriae participate in donor-strand-exchange of their N-terminal extensions (Nte), we re-oriented the Nte of CsnA to fit into the groove of the preceding subunit. Additional subunits were added with the known helical symmetry of CS20 fimbriae using UCSF Chimera, and checked for steric interference. This modeling can be seen in Fig. 4B-F. Upon inspection of the fit, a small loop was noted outside of the three-dimensional reconstruction (Fig. 4B lower white oval; Fig. 4E, black arrows). This loop is flanked by long loops on either side and therefore is expected to be inherently flexible. Using this flexibility, a better fit was achieved when the loop was moved upward and to the right (Fig. 4A gold ribbon, black arrows; Fig. 4B, upper white oval; Fig.4C, black arrow). Furthermore, residues 202-204, 206-208, and 210 (Fig. 4A blue ribbon) contained in the re-positioned alpha helix and adjoining loop have the ability to bond to the subunit above it (within residues 228-239). This movement as well as the repositioned Nte to accommodate donor-strand-exchange can be seen in Fig. 4A, C, and E.

For comparison, CsnA is shown in the CS20 fimbria map (Fig. 5A, green), CfaB is shown in the CFA/I fimbria map (Fig. 5A, blue), and PapA is shown in the P-fimbria map (Fig. 5A, pink). In addition, Fig. 5B shows surface views of these fimbriae at thresholds enclosing 50 % of the expected volume. Layer-to-layer interactions are indicated by solid black lines and defined as interactions between subunits from two adjacent layers are shown in the insets taken from within the black ovals drawn on each fimbria. For P-fimbriae these interactions are stronger than for CS20 or CFA/I, since each P-fimbrial subunit interacts with two subunits in an adjacent layer, as indicated by black and gray solid lines. Properties of these structures are shown in table S2; while FimA fitted into a Type 1 fimbria map is not available, the phylogenetic relationship of major pilins from a wide range of adhesion fimbriae is shown in Fig. S4 (Dereeper *et al.*, 2008). An isosurface of one subunit is shown for each of the maps. The CS20 subunit sits 2º from horizontal while the CFA/I subunit sits at -2º from horizontal, and the P-fimbriae subunit sits 28º from horizontal with respect to the helical axis. The degree of tilt for the PapA subunit reported here is dissimilar from our previous report of P-fimbriae (13º from horizontal; (Mu & Bullitt, 2006). This is not a change in the pilin position, but a change in our measurement procedure; previously, we measured by the center of mass, while in this study the subunit was measured by orientation of the bottom of the subunit with respect to the horizontal axis. This was done in order to have a more rigorous standard measurement technique by which to measure the other subunits.

**Discussion**

Pathogen colonization requires sustained bacterial binding, generally accomplished by essential adhesion fimbriae. These fimbriae function in diverse environments, requiring specialized biophysical properties adapted to their preferred target tissues. For example, unwinding of fimbriae can facilitate sustained binding by helping to reduce the load on the adhesin when bacteria are exposed to shear forces. We show in this work that CS20 fimbriae can be unwound up to eight times their initial length, similar to previously investigated helical fimbriae including UPEC-expressed Type 1, P, and F1C fimbriae; ETEC-expressed CFA/I, and respiratory-expressed Type 3 fimbriae (Mu & Bullitt, 2006, Mu *et al.*, 2008, Bullitt & Makowski, 1995, Jass *et al.*, 2004, Andersson *et al.*, 2007, Forero *et al.*, 2006, Chen *et al.*, 2011, Castelain *et al.*, 2011). Despite their common helical architecture, the structural details (e.g., Fig. 5 and Table S2), biomechanical properties (e.g., Table S1), amino acid identity and phylogenetic similarity (Table 1 and Fig. S4) of each fimbrial type vary. It is therefore of interest to assess these differences and determine to what extent they can be correlated to the preferred *in vivo* environment of the pathogen. Firstly, UPEC P and Type 1 fimbriae are commonly found on bacteria in cases of kidney or bladder infections, respectively (Andersson *et al.*, 2007). These fluid environments are complex, with boluses slowly transported from the kidney to the bladder, and then a high fluid velocity during expulsion of urine from the bladder. Secondly, where ETEC initiate infection in the intestinal ileum, peristaltic motion creates shear forces that include reversal wall shear forces and normal forces (Jeffrey *et al.*, 2003). Thirdly, in the respiratory tract, where Hib and Type 3 fimbriae attach, gaseous flushes (e.g., sneezes) can be severe, reaching velocities up to 50 m/s (100 mph; (Xie *et al.*, 2007)). From Table 1, it is clear that sequence similarity does not correlate directly with environment, as the CS20 major pilin has the lowest homology to the CFA/I major pilin (15% identity and 36% similarity), yet both are expressed on ETEC that initiate infection in the intestines. This supports the suggestion that CFA/I and CS20 may have emerged through convergent evolution (Sakellaris & Scott, 1998). Moreover, the CS20 major pilin shares 22-23% identity and 42% similarity with Type 1 and P-fimbriae, both of which colonize different epithelial tissues. Conversely, type 3 fimbriae on respiratory pathogens share 16% identity and 36% similarity with CS20. Thus, a simple amino acid comparison of the major structural pilin proteins of different fimbriae does not correlate with the pathophysiological niche of each *E. coli* strain.

The unwinding force for CS20 fimbriae is 8 pN higher than the force needed to unwind CFA/I fimbriae (Andersson *et al.*, 2012), 14 pN lower than the force needed for Type 1 and P-fimbriae, and more than 50 pN lower than the force needed to unwind Type 3 fimbriae. To understand these differences, the three-dimensional structures of P-fimbriae, CS20 and CFA/I fimbriae were compared. While no structural model of Type 3 fimbriae from *Klebsiella pneumoniae* is currently available, a low resolution 3D reconstruction of HibA fimbriae indicates a 3-stranded rope-like structure which is expected to eliminate its ability to unwind (Mu *et al.*, 2002). The layer-to-layer interactions are strongest for P-fimbriae, intermediary for CS20 fimbriae, and weak for CFA/I fimbriae. CFA/I and CS20 fimbriae are both critical colonization factors of enteric diseases, have weaker layer-to-layer interactions, and unwind at a lower force than adhesion fimbriae expressed by UPEC. Conversely, regardless of their different target environments, CS20 and P-fimbriae are homologous with respect to their primary sequence and share structural similarities. In particular, both structures display layer-to-layer interactions that are strongest along the left-handed, three-start helix.

Structural comparisons of CS20, P-fimbriae and CFA/I fimbriae showed that measurements on CS20 fimbriae reported here agree with previously published subunits/turn, filament length, and inner and outer diameters of other fimbriae (Mu *et al.*, 2008, Mu & Bullitt, 2006). CS20 fimbriae appear to have more intersubunit connections than CFA/I but fewer than P-fimbriae, as expected from force spectroscopy data that quantitated the force required for fimbrial unwinding. This range of interaction is seen clearly when reconstructions are thresholded to display only the strongest densities (Fig. S6). Homology modeled and energy minimized CsnA subunits fitted into the EM map suggest the presence of 23 subunit-subunit interactions of residues that share sequence similarity to PapA. While we speculate that the amino acid similarity between CsnA and PapA may be partly responsible for this suggested bonding between adjacent layers, an atomic resolution structure of CsnA from CS20 fimbriae is needed to test this hypothesis.

In addition to being more easily unwound, the maximum velocities at which no additional force is required for unwinding (corner velocity) are higher for ETEC fimbriae CS20 and CFA/I, as compared to UPEC Type 1 and P-fimbriae. When examined in more detail, the corner velocity appears to depend strongly on the pathophysiological environment of the bacteria. Type 1 fimbriae on UPEC bacteria that colonize the bladder have a very low corner velocity implying that they will extend slowly when exposed to tensile force that is higher than the unwinding force. To remain bound during rapid urine expulsion Type 1 fimbriae need to modulate the force on the FimH adhesin, which is of catch-bond type, so it rapidly reaches the optimal force to provide the longest lifetime. Thereafter they need to keep the force at a sustainable level, which is provided by unwinding. Thus, Type 1 fimbriae must not unwind too quickly in the bladder. The increased force required for unwinding allows them therefore to remain intact under short, rapid flow. P-fimbriae on bacteria that infect kidneys have a much higher corner velocity, indicating faster unwinding that results in a reduced drag force on the bacterium, as compared to Type 1 fimbriae (Zakrisson *et al.*, 2012). Presumably this property enhances sustained attachment to target tissues during the time periods of fluid flow in the kidneys. ETEC fimbriae have still higher corner velocities, likely so that the quick, strong forces in the intestines are absorbed by fast unwinding of the fimbriae.

These data, taken together, highlight the key role that environment plays in adaptations of fimbrial structure, organization, and optimized biophysical properties. Fimbriae that experience more consistent shear forces in the kidneys have additional layer-to-layer interactions as compared to fimbriae that must remain bound during short bursts of strong forces in the intestines. CFA/I fimbriae, with the lowest measured unwinding force, produce the most unwound fimbriae, as observed by electron microscopy. Surprisingly, however, despite significantly lower forces needed to unwind CS20 as compared to P-fimbriae, we observe many fewer unwound CS20 fimbriae under our experimental conditions. We suggest that the near-horizontal subunit orientation distributes shear forces more evenly across the interactions during fluid flow, as compared to the PapA subunit of P-fimbriae with its 28-degree tilt. From the observed biophysical and structural similarities and differences reported here, we conclude that pathophysiological environment plays a critical role in adaptations of fimbriae from diverse bacterial strains.

## Experimental Procedures

**Bacterial strains, growth conditions, and peptides**

WS7179A-2 is a wild type ETEC strain that expresses the heat-labile enterotoxin (LT) and heat-stable enterotoxin (STp), and serotypes as O17:H45. It was isolated from a young Egyptian child with diarrhea in a previously conducted prospective study of childhood diarrhea (Shaheen *et al.*, 2004) and provided for use here from a prior collection, with permission. The recombinant plasmid pRA101 was generated by insertion of the positive transcriptional regulator gene *cfaD* into a low-copy-number plasmid vector (Wang & Kushner, 1999). CS20 fimbriae were expressed from WS7179A-2/pRA101. Bacteria were grown on LB agar plates with 50 mg/ml kanamycin at 37° C for 24 h. Cultures were then restreaked on CFA plates and incubated at 37°C for another 24 h.

**DNA preparation and nucleotide sequence analysis**

For DNA sequence analysis, wild-type plasmid DNA was purified by a modified alkaline lysis procedure (plasmid midi kit; QIAGEN, Valencia, Calif.). Initial primers were designed from the published sequence of the CS20 major subunit (Valvatne *et al.*, 2004), and thereafter the complete sequence of both strands of the entire CS20 gene cluster was then generated by primer walking. Sequencing reactions were performed with Big Dye terminator enzyme mix (Applied Biosystems, Foster City, Calif.) in a Perkin-Elmer thermal cycler using standard conditions, and the reactions were analyzed on an ABI PRISM 3100 genetic analyzer (Applied Biosystems). Analysis of the derived sequences was performed with Sequencher, version 4.1 (Gene Codes Corporation, Ann Arbor, Mich.).

**Sample preparation and optical tweezers measurements**

Beads (9.5 µm, IDC, 2-10000) were diluted 1:500 in MilliQ water. A droplet of 10 µl was added to a coverslip #1 and placed in a 60°C oven for 1 h. This procedure immobilized the beads to the coverslip. 10 µl of a 0.01% poly-L-lysine (Sigma-Aldrich, Stockholm, Sweden) solution was thereafter added to functionalize the 9.5 µm beads. The cover slip was put in an incubator at 37°C for 60 minutes. Two ~5×15 mm strips of parafilm were cut and placed on the coverslip, and a new smaller coverslip was added on top forming a flow chamber. The flow chamber was placed on a heating plate ~80°C and by gently pressing the top coverslip a robust flow chamber was formed of ~100 µm height and with two open ends.

Prior to force spectroscopy measurements a colony from a CFA agar plate was harvested and diluted in 1 ml of PBS. 5 µl of suspended bacteria and 2.5 µm beads (Duke Scientific Corp., suspended in PBS 1:500), were injected into a chamber and the two ends of the chamber were sealed by vacuum grease.

Force spectroscopy measurements of individual fimbriae were performed by use of optical tweezers (OT) using methods and instrumentation previously described (Fällman *et al.*, 2004, Andersson *et al.*, 2008). In brief, the OT instrumentation is built around an inverted microscope (Olympus IX71, Olympus, Japan) with a high numerical aperture oil immersion objective (model: UplanFl 100X N.A. = 1.35; Olympus, Japan). A continuous Nd:YVO$_4$ laser (model: Millennia IR) that operates at 1064 nm, is used for trapping bacteria and probe beads. The position of a bead in the trap is monitored by projecting the refracted beam of a low power HeNe-laser onto a position sensitive detector. The stability of the setup was optimized to reduce drift and noise by measuring long time series and by using the Allan variance method described in (Andersson *et al.*, 2011). For DFS, the program that controlled the translation was set to unwind the fimbria for 1.5 µm between two defined positions at velocities of: 0.05, 0.15, 0.45, 1.35, 4.05, 12.15, 36.45 µm/s. Between each cycle the fimbria was rewound at 0.1 µm/s and the motion was paused for 1 s before the next pull was initiated. To determine the surface charge on CS20 fimbriae, bacteria were immobilized to 9.5 µm spheres according to the procedure described in the methods section. A small probe bead of either aldehyde/sulfate or amidine surface was trapped and brought into contact with an immobilized bacterium. If the bead did not interact with the fimbriae of the bacterium after several trials the experiment was reported as a failure, whereas if the bead attached and it was possible to unwind a fimbria the test was successful.

**Atomic force microscopy imaging**

AFM micrographs of bacterial cells were obtained using a procedure described earlier (Balsalobre *et al.*, 2003). Initially cells were suspended in 50 µl of filtered water, 10 µl of which was placed onto freshly cleaved ruby red mica (Goodfellow Cambridge Ltd., Cambridge). The cells were then incubated for 5 min and blotted dry before they were placed into a desiccator for a minimum of 2 h. Micrographs were collected with a Nanoscope V Multimode8 AFM equipment (Bruker software) using Bruker ScanAsyst mode with Bruker ScanAsyst-air probe oscillated at resonant frequency of 50-90 kHz.

**CS20 purification and EM Imaging**

*E. coli* expressing CS20 fimbriae were grown at 37°C overnight, pelleted and resuspended in PBS. The sample was then heated to 65°C for 25 minutes to extract the fimbriae, cells were removed by centrifugation at 10,000 g for 30 minutes and 0.24 g/ml ammonium sulfate was added to the supernatant and rocked at room temperature for at least 2 hours to precipitate the fimbria. Those fimbriae were then collected by centrifugation at 12,000 g and resuspended in PBS and the ammonium sulfate precipitation was repeated. The final preparation of resuspended fimbriae was then dialysed against TE buffer (10mM Tris, 0.1 mM ethylenediaminetetraacetic acid, pH 7.4).

For cryoelectron microscopy, 3-4 μl of sample fimbriae was applied to a Quantifoil® grid (Jena, Germany) then blotted with filter paper for 0.5-2 seconds before being rapidly plunge-frozen in liquid ethane by a Vitrobot (FEI, Oregon). Fimbriae were imaged with a JEOL2100 at NCMI with an accelerating voltage of 200kV, recorded on Kodak SO163 film and scanned on a CreoScitex scanner with a pixel size of 1.81 Å/pix. Images used for 3D reconstruction were taken at defocus values ranging from -0.8 to -3.5 μm, and corrections for phases of the contrast transfer function were done on a per-micrograph basis.

**STEM and Persistence Length Analysis**

PCMass software from Brookhaven National Lab (BNL) STEM facility (Wall, 2012) was used to determine mass/unit length from STEM images taken at BNL, using TMV as an internal calibration standard. Persistence length was calculated using a method described earlier (Li *et al.*, 2010).

**Molecular Modeling**

A homology model of CsnA was built using FimA as an initial reference with MODELLER software program (Sali & Blundell, 1993). The model was verified using the hydrophobic and surface potential information from UCSF Chimera (Pettersen *et al.*, 2004).

**Dynamic behavior of a fimbria, a theoretical description**

The quaternary structure of a fimbria is formed of subunits connected via layer-to-layer interactions that unwind when exposed to tensile stress. The unwinding and rewinding behavior, thereby the extension and contraction of fimbria, can be described by rate theory. As mentioned in ref. (Andersson *et al.*, 2006a) the effective bond opening rate of a single fimbria is given by the difference between the bond opening and closure rates. Therefore the unwinding velocity $\dot{L}$ of a helical fimbria under tensile force, $F_P$, can be expressed as the difference between the bond opening and closing rates times the bond opening length, $\Delta x_{AB}$

. Using the nomenclature defined by Zakrisson et al. (Zakrisson *et al.*, 2012), the corner velocity, $\dot{L}^*$, and the steady-state uncoiling force, $F_{SS}$, the unwinding velocity can be expressed as,

$$\dot{L}(F_P) = \dot{L}^* e^{(F_P - F_{SS})\Delta x_{AT}\beta} [1 - e^{-(F_P - F_{SS})\Delta x_{AB}\beta}], \tag{1}$$

where $\Delta x_{AT}$ is the bond length (from the ground state to the transition state), $\beta = 1/kT$, where $k$ is the Boltzmann's constant and $T$ is the temperature, and $\Delta x_{AB} = \Delta x_{AT} + \Delta x_{TB}$, where $\Delta x_{TB}$ is the distance from the transition state to the open state. The unknown entities, i.e., $\dot{L}^*$, $\Delta x_{AB}$, and $\Delta x_{AT}$, can thereby be derived by fitting Eq. (1) to the experimental data.

The dynamic behavior of the fimbriae that appears for high extension velocities ($\dot{L} > \dot{L}^*$) occurs when bond opening rate becomes significantly larger than the bond closing rate. Under these condition the rewinding rate can be neglected and the force in the system is given by a simplified model as,

$$F(\dot{L} > \dot{L}^*) = k_B T / \Delta x_{AT} \ln(\dot{L} / \dot{L}^{th}), \tag{2}$$

where $\dot{L}^{th}$ is the thermal extension velocity given by, $k_{AB}^{th} \Delta x_{AB}$, where $k_{AB}^{th}$ denotes the thermal bond opening rate (Andersson *et al.*, 2006a). This equation shows that the force required for unwinding increases as the extension velocity increases in a logarithmic manner, which can be referred as dynamic conditions. From Eq. (2) it is possible to derive the bond length by fitting to the linear part of a dynamic force-vs.-velocity diagram.

## **Acknowledgments**

We are grateful to Pawel Penczek for assistance with SPARX Helicon, the STEM facility at BNL, Justin Paluba and Emily Pontzer for technical assistance, National Center for Macromolecular Imaging (NCMI) at Baylor College of Medicine for data collection, and the Boston University Shared Computing Cluster managed by IS&T Computing Services for processing time. This work was supported by NIH (GM05722 and RR025434 to E.B.), the Swedish Research Council (621-2013-5379 to M.A.; 2010-3031 and 2012-4638 to B.E.U.), and the Carl Trygger foundation to M.A. The research was also supported by the U.S. Army Military Infectious Diseases Research Program Work Unit Number A0307 (to S.J.S.), and by the Henry M. Jackson Foundation for the Advancement of Military Medicine (S.J.S.). The views expressed in this article are those of the authors and do not necessarily reflect the official policy or position of the Department of the Navy, Department of Defense, nor the U.S. Government. S.J.S. is a military service member. This work was prepared as part of his official duties. Title 17 USC. §105 provides that 'Copyright protection under this title is not available for any work of the United States Government.' Title 17 U.S.C. §101 defines a U.S.

**References**

Andersson, M., O. Axner, F. Almqvist, B.E. Uhlin & E. Fällman, (2008) Physical Properties of Biopolymers Assessed by Optical Tweezers: Analysis of Folding and Refolding of Bacterial Pili. *ChemPhysChem* **9**: 221-235.

Andersson, M., O. Björnham, M. Svantesson, A. Badahdah, B.E. Uhlin & E. Bullitt, (2012) A Structural Basis for Sustained Bacterial Adhesion: Biomechanical Properties of CFA/I Pili. *J Mol Biol* **415**: 918-928.

Andersson, M., F. Czerwinski & O. LB, (2011) Optimizing active and passive calibration of optical tweezers. *J Opt* **13**.

Andersson, M., E. Fällman, B.E. Uhlin & O. Axner, (2006a) Dynamic Force Spectroscopy of E. coli P Pili. *Biophys J* **91**: 2717-2725.

Andersson, M., E. Fällman, B.E. Uhlin & O. Axner, (2006b) A Sticky Chain Model of the Elongation and Unfolding of Escherichia coli P Pili under Stress. *Biophys J* **90**: 1521-1534.

Andersson, M., B.E. Uhlin & E. Fällman, (2007) The Biomechanical Properties of E. coli Pili for Urinary Tract Attachment Reflect the Host Environment. *Biophys J* **93**: 3008-3014.

Balsalobre, C., J. Morschhäuser, J. Jass, J. Hacker & B.E. Uhlin, (2003) Transcriptional Analysis of the sfa Determinant Revealing Multiple mRNA Processing Events in the Biogenesis of S Fimbriae in Pathogenic Escherichia coli. *J Bacteriol* **185**: 620-629.

Bao, R., A. Fordyce, Y.-X. Chen, A. McVeigh, S.J. Savarino & D. Xia, (2014) Structure of CfaA Suggests a New Family of Chaperones Essential for Assembly of Class 5 Fimbriae. *PLoS Pathog* **10**: e1004316.

Behrmann, E., G. Tao, D.L. Stokes, E.H. Egelman, S. Raunser & P.A. Penczek, (2012) Real-space processing of helical filaments in SPARX. *J Struct Biol* **177**: 302-313.

Bullitt, E. & L. Makowski, (1995) Structural polymorphism of bacterial adhesion pili. *Nature* **373**: 164-167.

Bullitt, E. & L. Makowski, (1998) Bacterial adhesion pili are heterologous assemblies of similar subunits. *Biophys J* **74**: 623-632.

Castelain, M., S. Ehlers, J. Klinth, S. Lindberg, M. Andersson, B. Uhlin & O. Axner, (2011) Fast uncoiling kinetics of F1C pili expressed by uropathogenic Escherichia coli are revealed on a single pilus level using force-measuring optical tweezers. *Eur Biophys J* **40**: 305-316.

Castelain, M., A. Sjöström, E. Fällman, B. Uhlin & M. Andersson, (2010) Unfolding and refolding properties of S pili on extraintestinal pathogenic Escherichia coli. *Eur Biophys J* **39**: 1105-1115.

Chen, F.-J., C.-H. Chan, Y.-J. Huang, K.-L. Liu, H.-L. Peng, H.-Y. Chang, G.-G. Liou, T.-R. Yew, C.-H. Liu, K.Y. Hsu & L. Hsu, (2011) Structural and Mechanical Properties of Klebsiella pneumoniae Type 3 Fimbriae. *J Bacteriol* **193**: 1718-1725.

Choudhury, D., A. Thompson, V. Stojanoff, S. Langermann, J. Pinkner, S.J. Hultgren & S.D. Knight, (1999) X-ray structure of the FimC-FimH chaperone-adhesin complex from uropathogenic Escherichia coli. *Science* **285**: 1061-1066.

Dereeper, A., V. Guignon, G. Blanc, S. Audic, S. Buffet, F. Chevenet, J.-F. Dufayard, S. Guindon, V. Lefort, M. Lescot, J.-M. Claverie & O. Gascuel, (2008) Phylogeny.fr: robust phylogenetic analysis for the non-specialist. *Nucleic Acids Research* **36**: W465-W469.

Dodson, K.W., F. Jacob-Dubuisson, R.T. Striker & S.J. Hultgren, (1993) Outer-membrane PapC molecular usher discriminately recognizes periplasmic chaperone-pilus subunit complexes. *Proc Natl Acad Sci U S A* **90**: 3670-3674.

Fällman, E., S. Schedin, J. Jass, M. Andersson, B.E. Uhlin & O. Axner, (2004) Optical tweezers based force measurement system for quantitating binding interactions: system design and application for the study of bacterial adhesion. *Biosens Bioelectron* **19**: 1429-1437.

Forero, M., O. Yakovenko, E.V. Sokurenko, W.E. Thomas & V. Vogel, (2006) Uncoiling mechanics of Escherichia coli type I fimbriae are optimized for catch bonds. *PLoS Biol* **4**: e298.

Girardeau, J.P., Y. Bertin & I. Callebaut, (2000) Conserved structural features in class I major fimbrial subunits (Pilin) in gram-negative bacteria. Molecular basis of classification in seven subfamilies and identification of intrasubfamily sequence signature motifs which might be implicated in quaternary structure. *J Mol Evol* **50**: 424-442.

Hohn, M., G. Tang, G. Goodyear, P.R. Baldwin, Z. Huang, P.A. Penczek, C. Yang, R.M. Glaeser, P.D. Adams & S.J. Ludtke, (2007) SPARX, a new environment for Cryo-EM image processing. *J Struct Biol* **157**: 47-55.

Honarvar, S., B.K. Choi & D.M. Schifferli, (2003) Phase variation of the 987P-like CS18 fimbriae of human enterotoxigenic Escherichia coli is regulated by site-specific recombinases. *Mol Microbiol* **48**: 157-171.

Jass, J., S. Schedin, E. Fallman, J. Ohlsson, U.J. Nilsson, B.E. Uhlin & O. Axner, (2004) Physical Properties of Escherichia coli P Pili Measured by Optical Tweezers. *Biophys J*.

Jeffrey, B., H.S. Udaykumar & K.S. Schulze, (2003) Flow fields generated by peristaltic reflex in isolated guinea pig ileum: impact of contraction depth and shoulders. *Am J Physiol Gastrointest Liver Physiol* **285**: G907-G918.

Larkin, M.A., G. Blackshields, N.P. Brown, R. Chenna, P.A. McGettigan, H. McWilliam, F. Valentin, I.M. Wallace, A. Wilm, R. Lopez, J.D. Thompson, T.J. Gibson & D.G. Higgins, (2007) Clustal W and Clustal X version 2.0. *Bioinformatics* **23**: 2947-2948.

Li, X., W. Lehman & S. Fischer, (2010) The relationship between curvature, flexibility and persistence length in the tropomyosin coiled-coil. *J Struct Biol* **170**: 313-318.

Liu, L., H.L. Johnson, S. Cousens, J. Perin, S. Scott, J.E. Lawn, I. Rudan, H. Campbell, R. Cibulskis, M. Li, C. Mathers & R.E. Black, (2010) Global, regional, and national causes of child mortality: an updated systematic analysis for 2010 with time trends since 2000. *The Lancet* **379**: 2151-2161.

Low, D., Braaten, B., van der Woude, M., (1996) Fimbriae. In: *Escherichia coli* and *Salmonella*. Cellular and Molecular Biology. F.C. Neidhardt (ed). Washington, D.C.: ASM Press, pp. 146-157.

Lozano, R., M. Naghavi, K. Foreman, S. Lim, K. Shibuya, V. Aboyans, J. Abraham, T. Adair, R. Aggarwal, S.Y. Ahn, M.A. AlMazroa, M. Alvarado, H.R. Anderson, L.M. Anderson, K.G. Andrews, C. Atkinson, L.M. Baddour, S. Barker-Collo, D.H. Bartels, M.L. Bell, E.J. Benjamin, D. Bennett, K. Bhalla, B. Bikbov, A.B. Abdulhak, G. Birbeck, F. Blyth, I. Bolliger, S. Boufous, C. Bucello, M. Burch, P. Burney, J. Carapetis, H. Chen, D. Chou, S.S. Chugh, L.E. Coffeng, S.D. Colan, S. Colquhoun, K.E. Colson, J. Condon, M.D. Connor, L.T. Cooper, M. Corriere, M. Cortinovis, K.C. de Vaccaro, W. Couser, B.C. Cowie, M.H. Criqui, M. Cross, K.C. Dabhadkar, N. Dahodwala, D. De Leo, L. Degenhardt, A. Delossantos, J. Denenberg, D.C. Des Jarlais, S.D. Dharmaratne, E.R. Dorsey, T. Driscoll, H. Duber, B. Ebel, P.J. Erwin, P. Espindola, M. Ezzati, V. Feigin, A.D. Flaxman, M.H. Forouzanfar, F.G.R. Fowkes, R. Franklin, M. Fransen, M.K. Freeman, S.E. Gabriel, E. Gakidou, F. Gaspari, R.F. Gillum, D. Gonzalez-Medina, Y.A. Halasa, D. Haring, J.E. Harrison, R. Havmoeller, R.J. Hay, B. Hoen, P.J. Hotez, D. Hoy, K.H. Jacobsen, S.L. James, R. Jasrasaria, S. Jayaraman, N. Johns, G. Karthikeyan, N. Kassebaum, A. Keren, J.-P. Khoo, L.M. Knowlton, O. Kobusingye, A. Koranteng, R. Krishnamurthi, M. Lipnick, S.E. Lipshultz, *et al.*, (2012) Global and regional mortality from 235 causes of death for 20 age groups in 1990 and 2010: a systematic analysis for the Global Burden of Disease Study 2010. *The Lancet* **380**: 2095-2128.

Mu, X.Q. & E. Bullitt, (2006) Structure and assembly of P-pili: a protruding hinge region used for assembly of a bacterial adhesion filament. *Proc Natl Acad Sci U S A* **103**: 9861-9866.

Mu, X.Q., E.H. Egelman & E. Bullitt, (2002) Structure and function of Hib pili from Haemophilus influenzae type b. *J Bacteriol* **184**: 4868-4874.

Mu, X.Q., Z.G. Jiang & E. Bullitt, (2005) Localization of a critical interface for helical rod formation of bacterial adhesion P-pili. *J Mol Biol* **346**: 13-20.

Mu, X.Q., S.J. Savarino & E. Bullitt, (2008) The three-dimensional structure of CFA/I adhesion pili: traveler's diarrhea bacteria hang on by a spring. *J Mol Biol* **376**: 614-620.

Nada, R.A., H.I. Shaheen, S.B. Khalil, A. Mansour, N. El-Sayed, I. Touni, M. Weiner, A.W. Armstrong & J.D. Klena, (2011) Discovery and Phylogenetic Analysis of Novel Members of Class b Enterotoxigenic Escherichia coli Adhesive Fimbriae. *Journal of Clinical Microbiology* **49**: 1403-1410.

Nishiyama, M., R. Horst, O. Eidam, T. Herrmann, O. Ignatov, M. Vetsch, P. Bettendorff, I. Jelesarov, M.G. Grütter, K. Wüthrich, R. Glockshuber & G. Capitani, (2005) Structural basis of chaperone–subunit complex recognition by the type 1 pilus assembly platform FimD. *EMBO* **24**: 2075-2086.

Nishiyama, M., T. Ishikawa, H. Rechsteiner & R. Glockshuber, (2008) Reconstitution of Pilus Assembly Reveals a Bacterial Outer Membrane Catalyst. *Science* **320**: 376-379.

Ochoa, T.J., F.P. Rivera, M. Bernal, R. Meza, L. Ecker, A.I. Gil, D. Cepeda, S. Mosquito, E. Mercado, R.C. Maves, E.R. Hall, A.-M. Svennerholm, A. McVeigh, S. Savarino & C.F. Lanata, (2010) Detection of the CS20 colonization factor antigen in diffuse-adhering Escherichia coli strains. *FEMS Immunology & Medical Microbiology* **60**: 186-189.

Pettersen, E.F., T.D. Goddard, C.C. Huang, G.S. Couch, D.M. Greenblatt, E.C. Meng & T.E. Ferrin, (2004) UCSF Chimera--a visualization system for exploratory research and analysis. *J Comput Chem* **25**: 1605-1612.

Puorger, C., M. Vetsch, G. Wider & R. Glockshuber, (2011) Structure, Folding and Stability of FimA, the Main Structural Subunit of Type 1 Pili from Uropathogenic Escherichia coli Strains. *J Mol Biol* **412**: 520-535.

Sakellaris, H. & J.R. Scott, (1998) New tools in an old trade: CS1 pilus morphogenesis. *Mol Micro* **30**: 681-687.

Sali, A. & T.L. Blundell, (1993) Comparative protein modelling by satisfaction of spatial restraints. *J Mol Biol* **234**: 779-815.

Sali, A., L. Potterton, F. Yuan, H. van Vlijmen & M. Karplus, (1995) Evaluation of comparative protein modeling by MODELLER. *Proteins* **23**: 318-326.

Sauer, F., K. Fütterer, J.S. Pinkner, K. Dodson, S.J. Hultgren & G. Waksman, (1999) Structural basis of chaperone function and pilus biogenesis. *Science* **285**: 1058-1061.

Shaheen, H.I., S.B. Khalil, M.R. Rao, R. Abu Elyazeed, T.F. Wierzba, L.F. Peruski, S. Putnam, A. Navarro, B.Z. Morsy, A. Cravioto, J.D. Clemens, A.-M. Svennerholm & S.J. Savarino, (2004) Phenotypic Profiles of Enterotoxigenic Escherichia coli Associated with Early Childhood Diarrhea in Rural Egypt. *Journal of Clinical Microbiology* **42**: 5588-5595.

Soto, G.E. & S.J. Hultgren, (1999) Bacterial adhesins: common themes and variations in architecture and assembly. *J Bacteriol* **181**: 1059-1071.

Tacket, C.O., G. Losonsky, H. Link, Y. Hoang, P. Guesry, H. Hilpert & M.M. Levine, (1988) Protection by milk immunoglobulin concentrate against oral challenge with enterotoxigenic Escherichia coli. *N Engl J Med* **318**: 1240-1243.

Valvatne, H., H. Steinsland, H.M. Grewal, K. Molbak, J. Vuust & H. Sommerfelt, (2004) Identification and molecular characterization of the gene encoding coli surface antigen 20 of enterotoxigenic Escherichia coli. *FEMS Microbiol Lett* **239**: 131-138.

Verger, D., E. Bullitt, S.J. Hultgren & G. Waksman, (2007) Crystal structure of the P pilus rod subunit PapA. *PLoS Pathog* **3**: e73.

Wall, J.S., (2012) Brookhaven National Lab Scanning Transmission Electron Microscopy Facility. In. http://www.bnl.gov/biology/stem/default.asp, pp.

Wang, R.F. & S.R. Kushner, (1999) Construction of versatile low-copy-number vectors for cloning, sequencing and gene expression in Escherichia coli. *Gene* **100**: 195-199.

Xie, X., Y. Li, A.T.Y. Chwang, P.L. Ho & W.H. Seto, (2007) How far droplets can move in indoor environments – revisiting the Wells evaporation–falling curve. *Indoor Air* **17**: 211-225.

Zakrisson, J., K. Wiklund, O. Axner & M. Andersson, (2012) Helix-like biopolymers can act as dampers of force for bacteria in flows. *Eur Biophys J* **41**: 551-560.

# Figure Legends

**Figure 1. Images of CS20 fimbriae expressed by enterotoxigenic *Escherichia coli* (ETEC).** A) Representative AFM micrograph of WS7179A-2/pRA101 *E. coli* cell expressing CS20 fimbriae. Approximate lengths of these fimbriae are ~2μm. Scale bar is 0.5μm. B) Cryo-EM (TEM) micrograph of frozen-hydrated isolated CS20 fimbriae. The micrograph shows intact (blue arrow), unwound (yellow arrow), and a vertically oriented region showing a fimbria end-on (red arrow). Few unwound regions were found. Scale bar is 500Å.

**Figure 2. Force spectroscopy measurement of a single CS20 fimbria.** The black curve represents extension of a fimbria with a translation velocity of 0.05 µm/s whereas the blue curve represents contraction at the same speed. The force-extension response is composed of three force regions representing morphological changes of the fimbriae; Region I – elastic stretching of the shaft, Region II – unwinding of the shaft, Region III – elastic stretching of the subunits. An overlay of the curves showing moving averages, each including 100 data points, is illustrated by the red curves. CS20 fimbriae unwind at a force of 15 ± 1 pN.

**Figure 3. Dynamic force spectroscopy measurements performed on an individual CS20 fimbria.** A) Force-extension curves illustrating the unwinding force of the quaternary structure of CS20 fimbriae in region II at the given velocities, for a distance of 1.5 µm. The force and position data were recorded at 5 kHz. B) Each data point represents the average fimbriae unwinding force from 13 measurements at each velocity. The gray dashed line represents the steady-state unwinding force, i.e., 15 pN. The blue dashed curve represents a fit of the simplified model (refolding neglected; Eq. (2)) to the data taken at high extension velocity where the slope of the fit yields a bond length of 0.4 nm. The red dashed line represents a full fit to the rate equations, Eq (10) in ref (Zakrisson *et al.*, 2012), which, in turn, yields the quantities; $\dot{L}^*$, $\Delta x_{AT}$, and $k_{AB}^{th}$.

**Figure 4. Homology model of CsnA subunit fit into 3D helical reconstruction of CS20 fimbria.** A) The most energetically stable homology model of CsnA (purple) at 0° (left) and 90° counter-clockwise rotation (CCW; right). The Nte was then modified to fit into the groove of the adjacent subunit (C and E, labeled Nte) and CsnA residues 202-217 were shifted slightly for a better fit into the 3D reconstruction (black arrow). The resulting structure is shown in gold. Blue ribbons highlight residues 202-204, 206-208, and 210 that may form layer-to-layer bonds. B) 3D reconstruction of CS20 fimbriae from cryoEM data shows a helical filament with an outer diameter of 82 Å and an inner diameter of 33.5 Å at 10.3 Å resolution. The surface map has been made partially transparent in order to see the fit of CsnA monomers. White ovals highlight residues before (lower oval), and after (upper oval), shifting to better fit the map. C-F) 15 Å slices

of CS20 fimbria surface map fitted with CsnA. CsnA is shown before (E, black arrow) and after (C, black arrow) optimization. Scale bar is 50 Å.

**Figure 5. Subunit localizations and layer-to-layer interactions in CS20, CFA/I, and P fimbriae.** Surface structures of CS20 (green), CFA/I (blue) and P fimbriae (pink) are shown in panel A. CryoEM reconstructions of CS20 and P fimbriae, and a model CFA/I fimbria (x-ray structure of CfaB fitted into a lower resolution EM map) are shown. The original reconstruction of CFA/I is shown in Fig S6. Each fimbria includes its representative subunit (tan). The CS20 fimbrial subunit CsnA fits 2º from horizontal while the CFA/I fimbrial subunit CfaB fits -2º from horizontal, and the P-fimbriae subunit PapA fits in 28º from horizontal, all perpendicular with respect to the helical axis. Scale bar is 50 Å. Panel B shows the layer-to-layer interactions at thresholds that include 50 % of the expected volume for CS20, CFA/I and P fimbriae, respectively. Layer-to-layer interactions are shown as black and grey lines.

**Table legend**

**Table 1. Identity and similarity of the major pilins.** The percent identity, similarity, and the pathophysiological niche of fimbriae when compared to CsnA from CS20. These numbers were generated using ClustalW2.

| Pilin | Fimbria | Identity (%) | Similarity (%) | Environment |
|-------|---------|--------------|----------------|-------------|
| CfaB  | CFA/I   | 19.7         | 48.9           | GI          |
| CsnA  | CS20    | --           | --             | GI          |
| FasA  | 987P    | 55.5         | 82.7           | GI          |
| FimA  | Type 1  | 24.3         | 61.0           | U           |
| FocA  | F1C     | 24.9         | 58.2           | U           |
| PapA  | P       | 22.8         | 58.9           | U           |
| SfaA  | S       | 23.7         | 56.5           | U           |
| HifA  | Hib     | 24.1         | 53.7           | R           |
| MrkA  | Type 3  | 20.8         | 56.3           | R           |

**Figures**

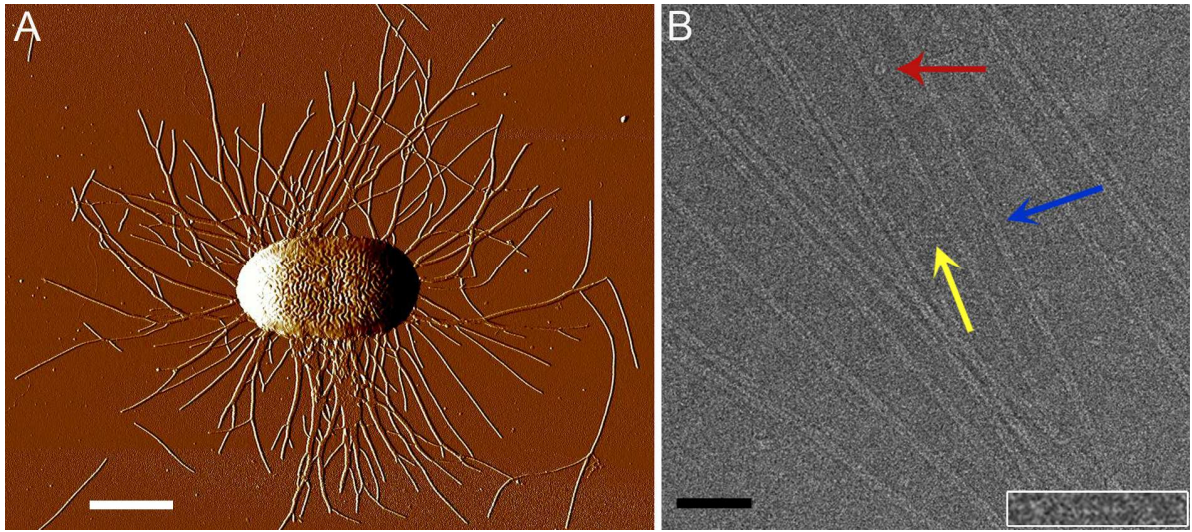

Figure 1

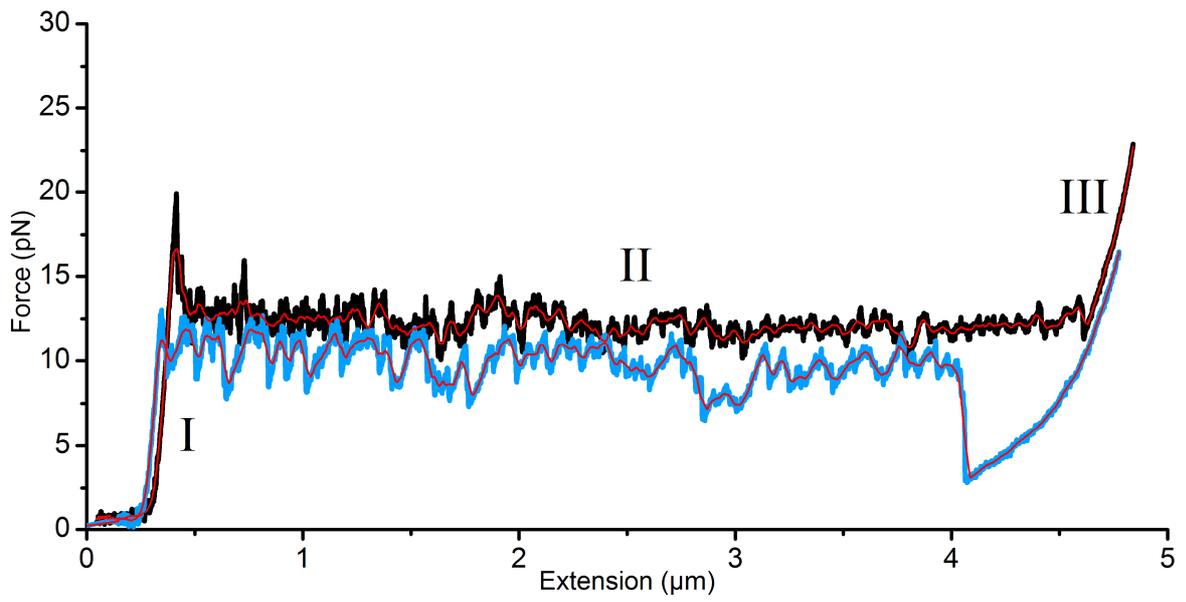

Figure 2

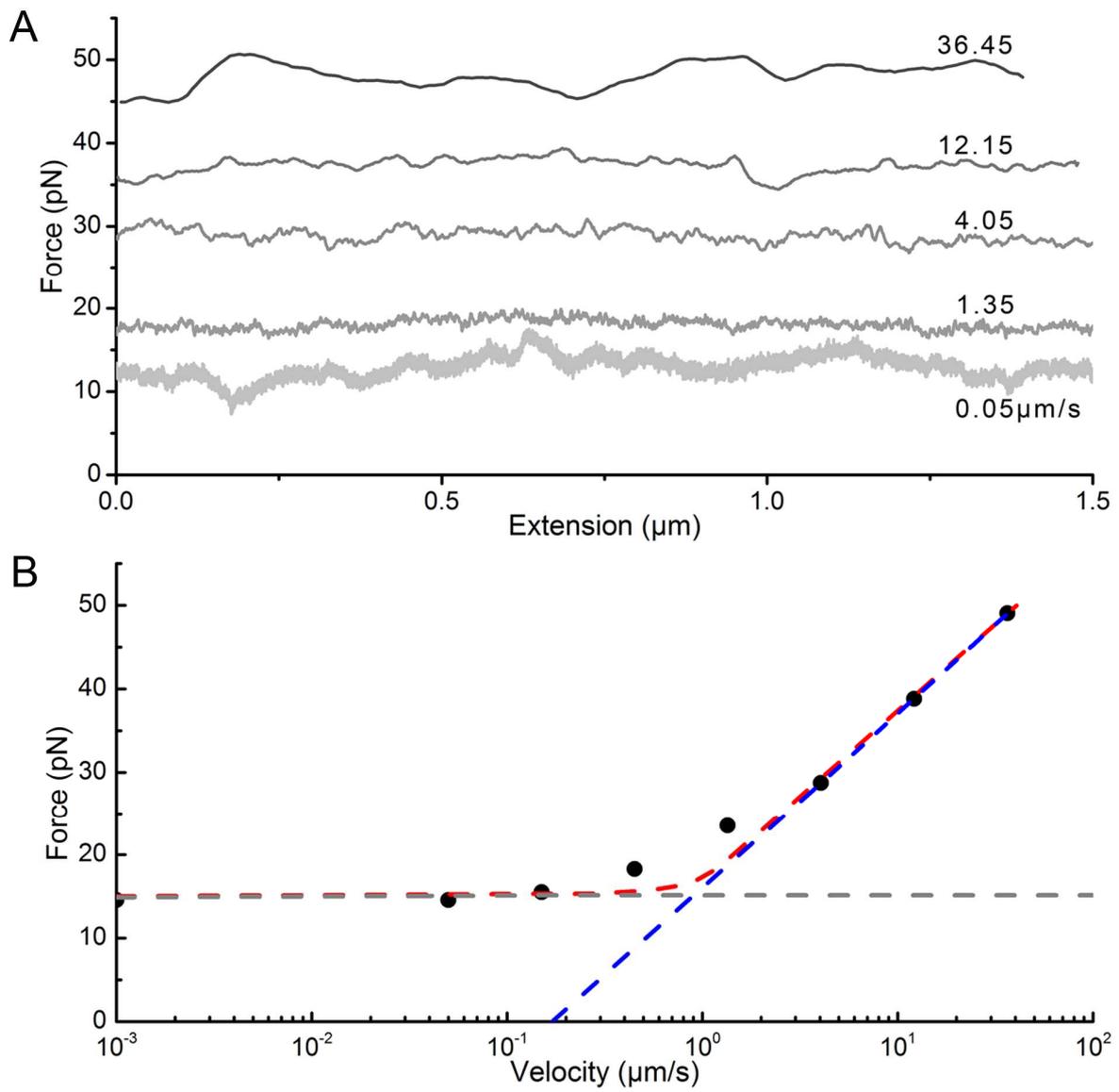

Figure 3

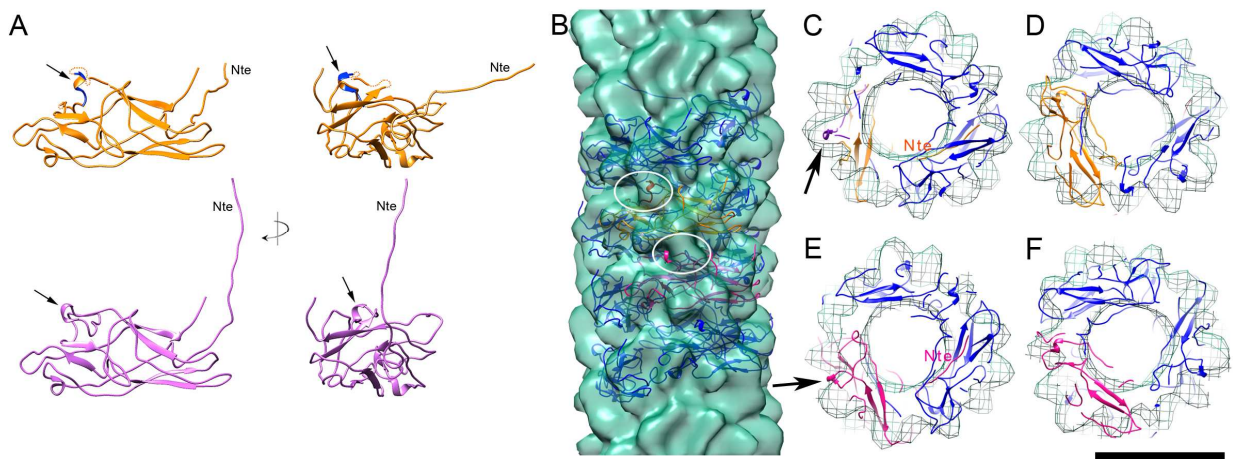

Figure 4

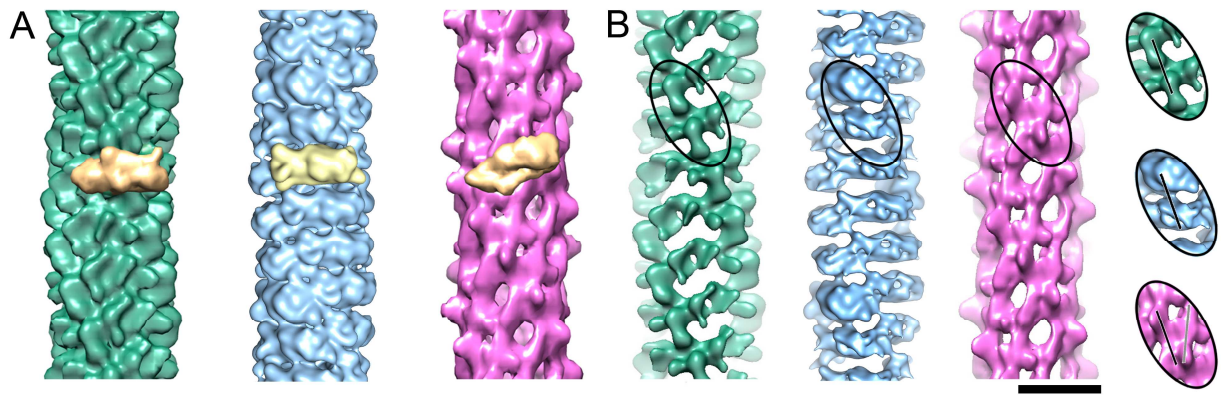

Figure 5